\documentclass[11pt,a4paper]{article}
\usepackage[T1]{fontenc}
\usepackage[utf8]{inputenc}
\usepackage{mathptmx}
\usepackage[margin=24mm]{geometry}
\usepackage{xcolor}
\usepackage{amsmath,amssymb}
\usepackage{graphicx}
\usepackage{booktabs,array,tabularx,ragged2e}
\usepackage{caption}
\usepackage{enumitem}
\usepackage{microtype}
\usepackage[numbers,sort&compress]{natbib}
\usepackage[hidelinks]{hyperref}
\usepackage{url}
\usepackage{placeins}

\newif\ifmarkup
\markupfalse           
\ifmarkup
  \newcommand{\blue}[1]{{\color{blue}#1}}
  
  \newcommand{\tblue}{\color{blue}}   
\else
  \newcommand{\blue}[1]{#1}
  
  \newcommand{\tblue}{}
\fi

\setlist{nosep}
\newcolumntype{L}[1]{>{\RaggedRight\arraybackslash}p{#1}}
\newcommand{\B}{\ensuremath{\mathcal{B}}}
\begin{document}

\begin{center}
{\LARGE\bfseries Builder, Defender, Breaker:\\[2pt]
\blue{Measurable Independence and Bounded Autonomy\\ When Generative Models Build, Defend and Test Software}}\\[10pt]

{\large Mohamed Chahine Ghanem$^{1,2,*}$}\\[6pt]
{\small
$^{1}$School of Computer Science and Mathematics, Keele University, Newcastle-under-Lyme, United Kingdom\\
$^{2}$Cybersecurity Institute, University of Liverpool, Liverpool, United Kingdom\\
$^{*}$Correspondence: \texttt{m.ghanem@keele.ac.uk}\quad {ORCID: 0000-0002-7067-7848}}\\[6pt]
{\small\itshape Short title: Builder, Defender, Breaker AI-led Security}
\end{center}

\vspace{4pt}
\noindent\textbf{Abstract.}
\blue{Generative models now write application code, harden and monitor it, and probe it for exploitable flaws, so that one family of models increasingly plays builder, defender and breaker at once. The prevailing view treats full autonomy as the natural end point of assistance. This article argues for a narrower and more defensible position than a blanket requirement for human oversight. We define the shared generative substrate as the set of upstream dependencies (training corpus, model family, alignment procedure, vendor, toolchain) that induce correlated errors, and we define independence as a measurable property of pairs of lifecycle roles rather than an attribute of any participant. A coincident-failure model in the tradition of Eckhardt and Lee shows why organisational independence, the proxy on which verification standards have relied, no longer implies statistical independence once roles share a substrate; it also shows that heterogeneous models, deterministic analysers and formal verification can restore independence for the fault classes they cover. What remains for humans is specific: authority over the specification against which every machine oracle is judged, accountability that current governance instruments do not permit to be delegated, and last-resort authority to halt. We translate this into an operational framework with five autonomy levels, three distinct human roles and five decision criteria (consequence, reversibility, time-criticality, verifiability and adversarial exposure), apply it to build, defend and test actions, and specify how independence and oversight effectiveness can be measured. The central claims are stated as testable hypotheses with a protocol for refuting them.}

\vspace{4pt}

\section{Three Hats, One Head}
Open a modern software pipeline and you are likely to find an AI model in at least three places. One model, or one family of models, drafts the application code. Another reviews that code, flags insecure patterns, and proposes hardening. A third generates test cases, fuzzes interfaces, and searches for exploitable flaws. The three functions correspond to the oldest division of labour in security practice: the builder who creates the system, the defender who protects it, and the breaker who tries to subvert it. What is new is that a single generative substrate is beginning to wear all three hats at the same time.

The prevailing narrative treats this as an unambiguous advance. If AI can assist a developer, why not let it write the whole module; if it can assist a defender, why not let it run the security operations centre; if it can assist a red team, why not let it hunt vulnerabilities unsupervised. Each step looks like a modest extension of the last, and each promises speed and scale that human teams cannot match. The trajectory points toward a lifecycle that builds, defends, and tests itself, with the human removed from the loop.

\blue{This article examines what is lost at that last step, and it tries to do so more carefully than the customary appeal to ``keeping a human in the loop''. The case for human oversight of consequential automation is already well made in trustworthy-AI guidance, in verification standards and in regulation; restating it would add little. Nor is it self-evident that the independence a security lifecycle needs must be supplied by a person. Heterogeneous models, deterministic analysers and formal methods are all candidate sources of independent judgement, and people have well-documented shared biases of their own. The contribution here is therefore not the principle but its operationalisation. We ask three questions that the principle leaves open. What, precisely, is lost when builder, defender and breaker share a generative substrate, and can the loss be measured rather than asserted? Which of the lost properties can be restored by technical means, and which cannot? And where, when and under what conditions does the answer require a human, as opposed to merely another independent mechanism?}

\blue{The evidence that automation helps is real, and nothing that follows disputes it. In the final round of DARPA's 2025 AI Cyber Challenge, cyber-reasoning systems operating under a strict no-human rule found 54 of the competition's 63 synthetic vulnerabilities (86 per cent), produced accepted patches for 68 per cent of those they found, and surfaced eighteen previously unknown flaws in real open-source code~\cite{darpaaixcc2025}. A framework that cannot accommodate that result is not a framework for practice. The framework set out below is designed to say where such systems should be allowed to run unsupervised, not merely where they should not.}

\blue{The article makes five contributions.
\begin{enumerate}[label=(C\arabic*),leftmargin=3.2em]
\item A definition of the \emph{shared generative substrate} and of \emph{independence} as a measurable property of pairs of lifecycle roles, together with a coincident-failure model, in the tradition of Eckhardt and Lee~\cite{eckhardt1985} and Littlewood and Miller~\cite{littlewood1989}, which explains why the organisational proxies for independence used by verification standards (different team, different vendor) cease to imply statistical independence once roles share a substrate (Section~\ref{sec:defs}).
\item A systematic appraisal of the alternatives to human involvement---heterogeneous model ensembles, deterministic tools, formal verification, policy gates and machine evaluators---that identifies which lost properties each restores, and a correspondingly narrow account of the three functions that remain for people: specification authority, accountability and last-resort halt (Section~\ref{sec:alternatives}).
\item An operational framework comprising five autonomy levels, three distinct human roles and five decision criteria, applied to concrete build, defend and test actions, with explicit conditions under which higher autonomy is justified (Section~\ref{sec:framework}).
\item Metrics and a protocol for measuring failure independence between roles and the effectiveness of human oversight, so that the framework's requirements can be audited rather than asserted (Section~\ref{sec:measure}).
\item An explicit treatment of counter-arguments and trade-offs, and a statement of the central claims as hypotheses with the observations that would refute them (Sections~\ref{sec:alternatives} and~\ref{sec:measure}).
\end{enumerate}
We are careful throughout to distinguish what has been measured from what has been argued. The proposition at the centre of the article---that roles drawn from one substrate fail together across build, defend and test---has not yet been measured directly across all three roles. Closely related measurements now exist for pairs of generated implementations and for models judging models~\cite{ron2026,nogueira2026,goel2025}, and they point the same way, but the three-role claim remains a hypothesis and is labelled as one.}

\blue{Section~\ref{sec:roof} reviews the evidence that each role is being automated, what the existing literature has already established, and how the present article differs from it, and then sets out the definitions and the formal model. Sections~\ref{sec:closed}--\ref{sec:accountability} analyse the four hazards of a closed loop (Figure~\ref{fig:loop}). Section~\ref{sec:alternatives} confronts the counter-arguments and derives the residual human contribution; Sections~\ref{sec:framework} and~\ref{sec:measure} give the framework and its measurement; Section~\ref{sec:discussion} discusses limitations, and Section~\ref{sec:conclusion} concludes.}

\section{A Lifecycle Under One Roof}\label{sec:roof}

\subsection{Building}
Code-generation assistants are now routine, and so is their central weakness: a model trained on the public corpus of human code reproduces the insecurity of that corpus. Pearce et al.\ prompted GitHub Copilot across scenarios drawn from the most dangerous common weakness enumerations and found that roughly forty per cent of the resulting programs contained exploitable vulnerabilities~\cite{pearce2022}. The pattern holds at scale: a formal analysis of some 331{,}000 C programs generated by nine state-of-the-art models found at least sixty-two per cent to be vulnerable, with the models differing only marginally from one another and exhibiting broadly the same coding errors~\cite{tihanyi2025}. These errors are not random. They track the vulnerability patterns latent in the training data, which means they are shared, to a first approximation, by every model trained on similar data. Worse, the humans nominally in charge do not compensate for them: in a controlled study, participants given an AI assistant wrote measurably less secure code \emph{and} were more likely to believe their code was secure~\cite{perry2023}.

\blue{The builder is also a target. Schuster et al.\ showed that a handful of crafted files inserted into the training or fine-tuning data of a neural code completer steer it towards insecure suggestions, and that the effect can be aimed at a particular repository~\cite{schuster2021}; Aghakhani et al.\ showed that the poison can be hidden in regions the model never reproduces verbatim, defeating signature-based cleansing~\cite{aghakhani2024}; and Carlini et al.\ showed that poisoning the web-scale datasets from which such models are trained is practical at modest cost~\cite{carlini2024}. A preprint from Hubinger et al.\ reports that a model trained to insert exploitable bugs under a trigger condition retained the behaviour through subsequent safety training~\cite{hubinger2024}; the result awaits peer review but is consistent with the rest. The relevance for what follows is that anything shared by builder, defender and breaker is shared as an attack surface too.}

\subsection{Defending}
On the defensive side, AI now drives intrusion detection, alert triage, malware classification, and configuration hardening. These systems learn a statistical picture of ``normal'' and flag departures from it. That picture is only ever as complete as the data behind it, and a decade of work catalogued by NIST's adversarial-machine-learning taxonomy shows that it can be evaded, poisoned, or misdirected by an adversary who understands how it was built~\cite{biggio2018,nistaml2025}. \blue{Where the defender is a language model asked to reason about code, its competence at the task is itself in question. Ding et al.\ found that a code model scoring 68 per cent F1 on a widely used vulnerability benchmark fell to about 3 per cent on the same task once duplicated and mislabelled samples were removed, with frontier models performing close to chance~\cite{ding2025}; Ullah et al.\ found that eight models gave non-deterministic and unfaithful vulnerability judgements, and that trivial renaming or the addition of unrelated library functions flipped GPT-4's answer in roughly one case in six and one in four respectively~\cite{ullah2024}. A defender that has internalised the same distribution as the code it protects will, on the model set out in Section~\ref{sec:defs}, tend to be blind to the same cases the builder was blind to; a defender whose judgement is this fragile offers little independent assurance even where its blind spots differ.}

\subsection{Breaking}
Offensive testing has been automated the longest, from fuzzers to symbolic-execution engines, and the frontier is now fully autonomous discovery and repair. DARPA's 2016 Cyber Grand Challenge staged the first all-machine hacking tournament, in which cyber-reasoning systems found, exploited, and patched flaws in seconds with no human in the loop~\cite{darpacgc2016}. \blue{Its successor, the 2025 AI Cyber Challenge, extended the ambition to real-world open-source software: with human interaction prohibited, the finalist systems identified 54 of the competition's 63 synthetic vulnerabilities (86 per cent), produced accepted patches for 68 per cent of those they found, and independently discovered eighteen genuine zero-day flaws, eleven of which they also patched, at an average cost of about \$152 per competition task~\cite{darpaaixcc2025}. The frontier is real but still bounded. PentestGPT, evaluated at USENIX Security, markedly improved sub-task completion over a bare model on penetration-testing benchmarks yet remained a human-in-the-loop tool~\cite{deng2024}; on Cybench's forty professional capture-the-flag tasks, unguided agents solved only those whose first human solve time in competition had been eleven minutes or less~\cite{zhang2025cybench}.} These programmes are genuine achievements. They also make the convergence concrete: the tools that test a system are now the same kind of system being tested.

Taken together, these three developments describe a lifecycle increasingly housed under one roof and, more consequentially, built from one kind of material. That shared material is the source of the problems that follow.

\subsection{What is already established, and what this article adds}\label{sec:established}
\blue{Three bodies of knowledge bear directly on the argument, and it is important to be exact about what each has settled. Separation of privilege is one of Saltzer and Schroeder's design principles~\cite{saltzer1975}, and independent verification and validation is codified in IEEE Std 1012, which defines independence along three axes---technical, managerial and financial---and characterises independent verification and validation as that performed by an organisation possessing all three forms of independence from the developer~\cite{ieee1012}. Software fault tolerance established, first theoretically and then empirically, that independently developed versions of the same specification do not fail independently: Eckhardt and Lee modelled the effect as a consequence of inputs that are difficult for everyone~\cite{eckhardt1985}, Knight and Leveson measured it in 27 independently written programs~\cite{knight1986}, and Littlewood and Miller showed that forced diversity of method can, in principle, do better than independence~\cite{littlewood1989}; Littlewood and Strigini carried the modelling into security~\cite{littlewood2004}. The monoculture argument is likewise mature: Geer et al.\ made it for operating systems in 2003~\cite{geer2003}, and Bommasani et al.\ made it for shared machine-learning models~\cite{bommasani2022}. Most recently, correlated error among language models has begun to be measured directly. Panickssery et al.\ found that model evaluators recognise and favour their own generations~\cite{panickssery2024}; Goel et al.\ introduced a chance-adjusted measure of agreement on mistakes and found that model errors are becoming more correlated as capability rises, which they argue undermines the use of models to oversee models~\cite{goel2025}; and two 2026 preprints have replicated the Knight--Leveson experiment with coding agents and found substantial common-mode failure~\cite{ron2026,nogueira2026}.}

\blue{Against that background the present article claims three things that are not, to our knowledge, established elsewhere. First, the independence proxies on which IEEE 1012 and its descendants rely are proxies for a statistical property, and a shared generative substrate breaks the link between proxy and property: two organisations that are technically, managerially and financially independent but license the same model, or fine-tune from the same base, or train on the same public corpus, satisfy the standard and share the blind spots. Independence must therefore be measured on the pair, not inferred from the organisation chart, and Section~\ref{sec:defs} says how. Second, the question ``can the independence be supplied by something other than a person?'' has a differentiated answer: yes for the fault classes a heterogeneous or deterministic oracle covers, no for the adjudication of what the specification ought to require, for accountability, and for halting a loop whose every internal check may be compromised. Third, the requirement for human involvement, so bounded, can be written as an operational framework whose conditions are auditable.}

\subsection{Definitions: substrate, independence and the coincident-failure model}\label{sec:defs}
\blue{\textbf{Shared generative substrate.} By the \emph{generative substrate} of a lifecycle role we mean the set of upstream dependencies that shape its error distribution: the pre-training corpus, the model family and its weights lineage (including any common base from which a fine-tune descends), the training and alignment procedure, the vendor's data and evaluation practices, and the toolchain and prompts through which the model is deployed. Two roles \emph{share} a substrate to the extent that these dependencies overlap. Sharing is graded, not binary, and Table~\ref{tab:dimensions} decomposes it into six dimensions with a note on what each induces and how it can be tested. The decomposition responds to a fair objection: ``a different model, training set or vendor'' is too coarse a prescription to act on, because two vendors' models trained on overlapping public corpora share dimension~D1 even though they differ on D3.}

\blue{\textbf{Independence as a property of pairs.} Let $X$ denote the space of situations a lifecycle role must handle---a specification to implement, an input to classify, a program to test---drawn according to the operational distribution $Q$. Following Eckhardt and Lee~\cite{eckhardt1985}, associate with each role a difficulty function: $b(x)$, the probability that the builder emits a flaw of a given class when handling $x$, and $t(x)$, the probability that the verifier (defender or tester) fails to detect that flaw when it is present. If the two roles are conditionally independent given $x$, a flaw of that class escapes the pair with probability $P_{\mathrm{esc}}=\mathbb{E}_Q[\,b(X)\,t(X)\,]$. The assumption that verification standards implicitly make is stronger: it treats the escape probability as $\mathbb{E}[b]\,\mathbb{E}[t]$, the product of the marginal rates. The difference is a covariance:
\begin{equation}\label{eq:cov}
P_{\mathrm{esc}} \;=\; \mathbb{E}_Q[b(X)]\;\mathbb{E}_Q[t(X)] \;+\; \operatorname{Cov}_Q\!\big(b(X),\,t(X)\big).
\end{equation}
Eckhardt and Lee's insight was that the covariance is positive even for independently developed versions, because some inputs are hard for everyone: if a common latent difficulty drives both functions, so that $b$ and $t$ are both non-decreasing in it, then $\operatorname{Cov}(b,t)\ge 0$ by Chebyshev's integral inequality, with equality only in degenerate cases, for instance when one role's difficulty does not vary across $X$. Knight and Leveson observed exactly this in human teams~\cite{knight1986}. A shared generative substrate strengthens the mechanism. Builder and verifier are not merely exposed to a common difficulty in the problem; they share the prior that determines what counts as difficult. A flaw pattern the substrate is disposed to emit is one to which it assigns low probability of being wrong, and is therefore one it is disposed not to flag. The covariance term in~\eqref{eq:cov} is the formal content of the phrase ``common blind-spot set'', and we write \B{} for the region of $X$ in which both $b$ and $t$ are high.}

\blue{Two consequences follow. First, independence is a property of the pair $(b,t)$ under $Q$, not of either participant: the same verifier may be independent of one builder and correlated with another, and a person who reads the machine's verdict before forming a view has a $t$ that is no longer independent of $b$ (Section~\ref{sec:humansbiased}). Second, the covariance can be estimated. On a probe set of situations with known flaws, the \emph{coincident-failure ratio}
\begin{equation}\label{eq:rho}
\rho \;=\; \frac{\hat P_{\mathrm{esc}}}{\hat P_{\mathrm{emit}}\;\hat P_{\mathrm{miss}}}
\end{equation}
compares the observed escape rate with the rate independence would predict, where $\hat P_{\mathrm{emit}}$ is the builder's flaw rate on the probe set and $\hat P_{\mathrm{miss}}$ is the verifier's miss rate on flawed items seeded independently of the builder. Equivalently, $\rho$ is the ratio of the verifier's miss rate on the builder's own flaws to its miss rate on flaws seeded independently: a verifier that misses the builder's flaws more often than it misses anyone else's is correlated with the builder. A value of $\rho=1$ indicates independence, $\rho>1$ correlated failure, and $\rho<1$ the ``better than independent'' regime that Littlewood and Miller showed forced diversity can achieve~\cite{littlewood1989}. Goel et al.'s chance-adjusted probability of agreement on mistakes is a closely related statistic for a pair of classifiers~\cite{goel2025}, and Nogueira et al.\ have proposed a methodology for assessing behavioural failure independence among generated implementations~\cite{nogueira2026}. The reason for writing the quantity down is that the argument of this article rests on claims about $\rho$, and claims about $\rho$ can be tested (Section~\ref{sec:measure}).}

\begin{table}[!htbp]
\centering\small
{\tblue
\caption{Dimensions along which two lifecycle roles may share a generative substrate. ``Different vendor'' addresses D3 and part of D2 only; D1 is typically shared across vendors, and D5--D6 are where non-generative oracles differ in kind.}
\label{tab:dimensions}
\begin{tabularx}{\textwidth}{@{}L{2.4cm} L{4.6cm} L{4.6cm} X@{}}
\toprule
\textbf{Dimension} & \textbf{What is shared} & \textbf{What sharing induces} & \textbf{How to test} \\
\midrule
D1 Data provenance & Overlapping pre-training or fine-tuning corpora (public code, vulnerability databases, documentation) & Common priors about ``normal'' code; common exposure to poisoned data~\cite{carlini2024,schuster2021} & Corpus-overlap disclosure; poisoning canaries planted in public sources \\
D2 Model lineage & Same base weights, distillation from a common teacher, same architecture family & Transferable adversarial inputs~\cite{demontis2019}; near-identical error profiles~\cite{tihanyi2025} & $\rho$ on a probe set; agreement-on-mistakes statistics~\cite{goel2025} \\
D3 Training and alignment procedure, vendor & Same reinforcement and safety-training data, same evaluation suites & Self-preference when judging~\cite{panickssery2024}; common refusal and hardening habits & Cross-vendor $\rho$; blind evaluation of authorship-masked outputs \\
D4 Toolchain and context & Same agent harness, system prompts, retrieval sources, scanners & Prompt-injection and configuration flaws replicated in every role & Configuration audit; injection canaries \\
D5 Verification method & Learned (statistical) versus deterministic (analysers, sanitisers, fuzzers) versus formal (proofs) & Learned verifiers share \B{} with a learned builder; deterministic and formal oracles do not, for the classes they cover & Per-class $\rho$; coverage mapping of deterministic oracles to fault classes \\
D6 Epistemic basis of the oracle & What the check is against: the model's own expectation, a written specification, or the requirement the system exists to meet & Specification-level errors pass every artefact-level check~\cite{barr2015,ron2026} & Specification review by the requirement owner; requirement-level canaries \\
\bottomrule
\end{tabularx}}
\end{table}

\FloatBarrier
\section{The Closed Loop: Correlated Failure and the Verifier}\label{sec:closed}

\begin{figure}[t]
\centering
\includegraphics[width=0.95\linewidth]{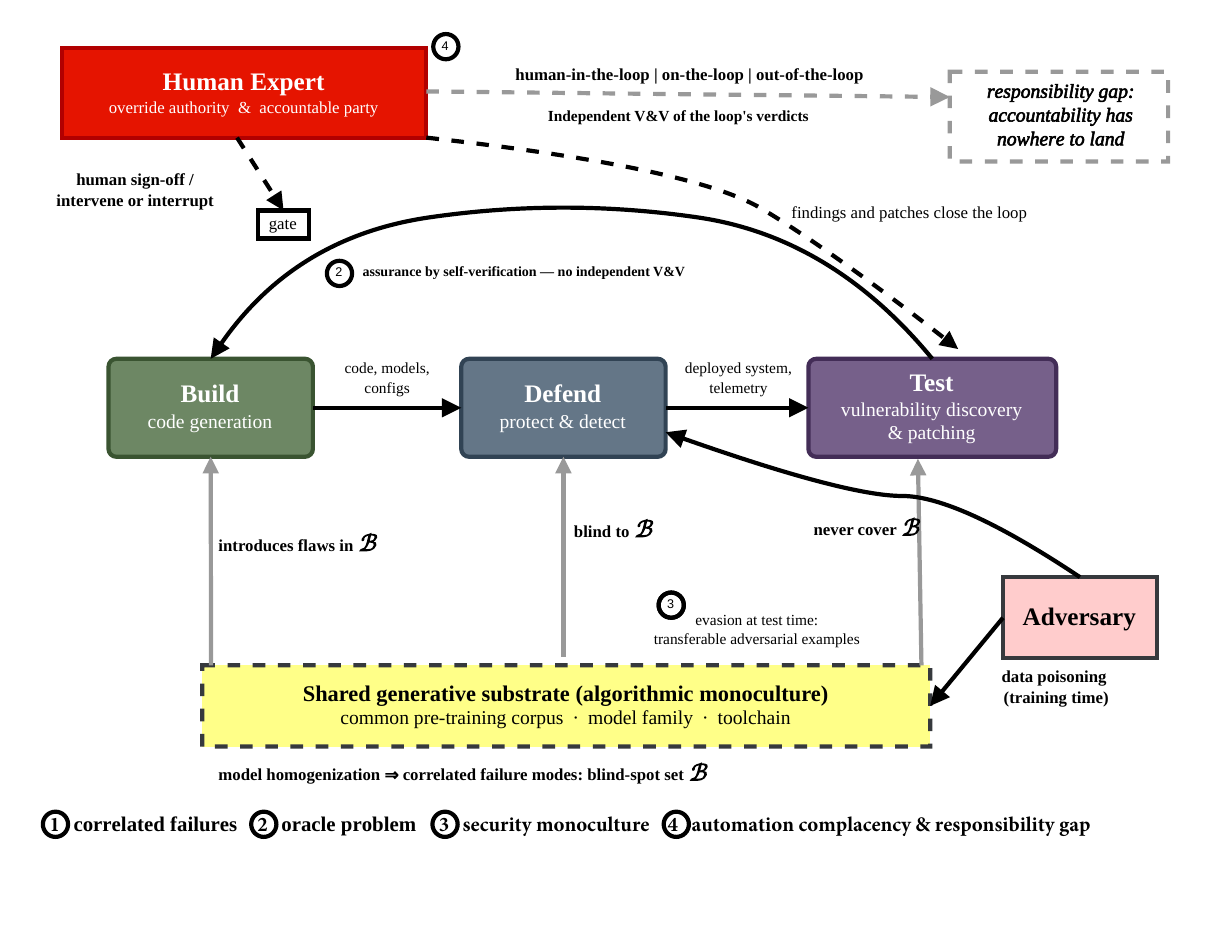}
\caption{\blue{The fully autonomous security lifecycle as a closed loop. Build, Defend and Test exchange artefacts along the solid path; findings and patches return to Build, and the loop's assurance rests on its own verdicts with no independent verification and validation (hazard~2). All three stages draw on one generative substrate whose shared priors define a common blind-spot set \B: flaws in \B{} are introduced by the builder, missed by the defender and never covered by the tester (hazard~1). The adversary faces a uniform, predictable target and can act on it at training time by poisoning and at test time through transferable adversarial examples (hazard~3). The human expert supplies what the loop does not generate for itself---independent verification of the loop's verdicts, sign-off on consequential actions and an accountable party---and under full autonomy drifts from in the loop to on the loop to out of it, leaving a responsibility gap (hazard~4). Sections~\ref{sec:closed}, \ref{sec:adversary} and~\ref{sec:accountability} analyse hazards~1--2, 3 and~4 respectively; Sections~\ref{sec:alternatives}--\ref{sec:framework} set out which mechanisms restore each property and where a person is required.}}
\label{fig:loop}
\end{figure}

\subsection{Correlated failure and the monoculture problem}
A security lifecycle works because its stages fail independently. The developer overlooks a flaw; the reviewer catches it; if the reviewer misses it too, the tester's exploit surfaces it. The protection comes not from any one stage being perfect but from the stages being unlikely to fail in the same place at the same time. Independence is the resource that defence in depth spends. Nor can it simply be assumed wherever redundancy exists: when Knight and Leveson tested the premise behind N-version programming, independently written implementations of a common specification failed on the same inputs far more often than the independence assumption allowed, because people trained alike make similar mistakes in the same hard places~\cite{knight1986}.

A homogeneous AI pipeline reproduces that experiment under \blue{less favourable} conditions. The builder, the defender, and the breaker are no longer merely educated alike; they are drawn from the same generative distribution, and their failures correlate accordingly (Figure~\ref{fig:loop}). A vulnerability that a model is disposed not to generate is, for the same reasons, one it is disposed not to detect and not to test for. The blind spot is not caught downstream because every stage downstream shares it. This is the security instance of algorithmic monoculture: when many decision-makers rely on the same underlying model, their outcomes homogenise, and errors that would have been idiosyncratic become systemic~\cite{bommasani2022}. The mechanism is not speculative. Demontis et al.\ showed that adversarial examples optimised against one model transfer to and defeat different models to the extent that those models share structure, with transfer success reaching the point of total misclassification and predictable from the alignment of their gradients~\cite{demontis2019}.

\blue{The direct evidence is now more than an analogy. Ron, Baudry and Monperrus replicated Knight and Leveson's experiment with 48 implementations of the original Launch Interceptor specification generated by five coding agents across 23 models and three languages, and found substantial common-mode failure concentrated where the specification was hard or ambiguous~\cite{ron2026}. Nogueira et al., across 224 problems, twelve models, five languages and three prompting strategies, found that implementations from different models fail on the same tests far more often than independence predicts, and that combining five of them by majority vote captured less than half of the reliability gain that independence would have allowed~\cite{nogueira2026}. Both studies also found that diversity helps, a point to which Section~\ref{sec:ensembles} returns; neither found anything close to independence; and both are preprints whose findings await peer review. Goel et al.\ add the observation most relevant to a self-verifying lifecycle: across many model pairs, errors grow more correlated as capability rises, and a model used as a judge is biased towards models that resemble it~\cite{goel2025}. Shared provenance is shared fragility. In an all-AI lifecycle it produces a class of defect that no stage is well placed to see.}

\subsection{Who verifies the verifier?}
Verification is meaningful only against an oracle independent of the thing verified; supplying such an oracle is the central and still open problem of software testing~\cite{barr2015}. A test suite written by the same process that wrote the code inherits the author's assumptions about what the code is supposed to do; it will confirm that the program does what the model expected, not that it does what the world requires. This is the machine analogue of marking one's own homework, and the concern is empirical rather than abstract. Self-preference has been measured directly: large-language-model evaluators recognise their own generations and score them more favourably than human annotators do, and the bias strengthens with the model's ability to recognise its own output~\cite{panickssery2024}. The human side of the arrangement fares no better. The one direct measurement we have of humans supervising an AI author found the opposite of a safeguard: participants with an assistant produced less secure code yet reported higher confidence in it~\cite{perry2023}.

\blue{A lifecycle that lets the substrate certify its own output has no oracle outside the substrate for the classes of flaw in \B. This is not the same as saying it has no independent oracle at all. A deterministic analyser, a sanitiser, a fuzzer with a crash oracle or a proof checker judges against something other than the model's expectation, and Section~\ref{sec:deterministic} gives these mechanisms their due. The point is narrower. Every such oracle checks an artefact against a stated property or specification; none can adjudicate whether the specification is what the world requires. The oracle problem, as Barr et al.\ frame it, bottoms out in that question~\cite{barr2015}, and the common-mode failures that Ron et al.\ traced to hard or ambiguous specification clauses are its signature~\cite{ron2026}. It is the gap that trustworthy-AI guidance addresses by requiring independent measurement and human review rather than a system's report on itself~\cite{nistrmf2023}.}

\FloatBarrier
\section{When Automation Meets the Adversary}\label{sec:adversary}
Security is adversarial, which distinguishes it from most other settings where AI is deployed. A recommender that makes a mistake faces a user who shrugs; a defence that makes a mistake faces an opponent actively looking for exactly that mistake, who will reuse it. Full autonomy interacts badly with this fact in three ways.

First, uniformity is exploitable. A pipeline built from one model family behaves predictably, and predictability is precisely what an attacker wants. Adversarial machine learning has spent a decade showing that learned detectors can be evaded by inputs crafted at test time and corrupted by data introduced at training time~\cite{biggio2018}; the attacks are not confined to the feature space but can be realised as working malware that a classifier passes as benign~\cite{pierazzi2020}. A monoculture magnifies the payoff, because such attacks transfer across models that share structure~\cite{demontis2019}: an evasion that works against one instance works against its siblings, and a homogeneous lifecycle offers no diverse second opinion to catch the manipulation. \blue{The training-time attacks surveyed in Section~\ref{sec:roof} compound this: a poisoned corpus is inherited by every role trained on it, so a single act of poisoning can corrupt the builder's habits, the defender's notion of normal and the tester's coverage at once~\cite{schuster2021,aghakhani2024,carlini2024}. Geer et al.\ made the same structural point about operating-system monoculture two decades ago: one flaw, all instances~\cite{geer2003}.}

Second, the offensive and defensive tools are the same tools. An autonomous system that can find and exploit vulnerabilities for the purpose of testing is, by description, an autonomous system that can find and exploit vulnerabilities. The dual use is no longer hypothetical: given a vulnerability description, an LLM agent was reported to autonomously exploit a majority of a set of real one-day flaws end to end, several of them rated critical~\cite{fang2024oneday}. When such a system runs without a human able to halt it, no one is positioned to notice that testing has shaded into damage, or that the tool has been redirected against a live target.

Third, autonomous judgment is brittle where it matters most, and its own reported successes reveal the limit. The same agent that exploited most flaws when handed a written description managed only a small fraction once that description was removed~\cite{fang2024oneday}: the competence is real but leans heavily on human-curated context. More generally, reasoning models can degrade sharply as problem complexity rises, with accuracy falling away past a threshold even as the model continues to emit confident output; the precise interpretation of this \blue{result, reported in a preprint, is contested,} but the fragility it exposes is not~\cite{shojaee2025}. The Cyber Grand Challenge supplies the parable. The winning autonomous system dominated the other machines and was then invited to the human Capture-the-Flag tournament at DEF~CON, where it finished last~\cite{darpacgc2016}. \blue{A decade of progress has narrowed that gap without closing it: on Cybench, unguided agents still solved only the tasks that competing human teams had first solved within minutes~\cite{zhang2025cybench}. Machine-speed autonomy won the contest that machines were built for and lost when it faced adversaries who could reason in ways it could not yet match.}

\section{Drift and the Accountability Vacuum}\label{sec:accountability}

\subsection{The vanishing human and machine-speed drift}
Even where a human is nominally retained, autonomy tends to push them out. The progression is familiar from other safety-critical domains: the human moves from \emph{in} the loop, approving each action, to \emph{on} the loop, monitoring a stream of actions, to \emph{out} of the loop, receiving a summary after the fact. Two forces drive the drift. The first is human: sustained oversight of a system that is usually right induces automation complacency and automation bias, an effect that Parasuraman and Manzey found in experts as well as novices and that resists training and instruction~\cite{parasuraman2010}. The nominal overseer stops genuinely checking. The second is machine speed: when a cyber-reasoning system patches a flaw in seconds, a human approval step is not a safeguard but a bottleneck, and it will be removed as one. The moment the loop \emph{can} close, both the incentive and the psychology push to close it. \blue{Bainbridge identified the further irony four decades ago: the more reliable the automation, the rarer and more demanding the occasions on which the human must take over, and the less practised the human is when they arrive~\cite{bainbridge1983}. Section~\ref{sec:framework} treats this not as an argument for more human involvement everywhere, which would merely accelerate the complacency, but for placing involvement where it can be exercised competently.}

\subsection{Where responsibility lands}
Suppose the closed loop nonetheless fails: an autonomous lifecycle ships a vulnerability, a defence poisons itself, a test escalates into an outage. Who is responsible? The question is not rhetorical, because accountability is \blue{a property that current governance instruments do not permit to be} delegated to a machine. A fully autonomous pipeline is the limit of a long-recognised erosion\blue{, which Matthias named the responsibility gap: as control over a learning system's behaviour passes from its designer to the system's own training, the conditions under which a person can fairly be held responsible for its actions are progressively removed~\cite{matthias2004}. In the closed loop} the code, the review, and the test all issue from a process with no principal, so the answerability that ought to attach to a decision has nowhere to land. It is precisely this that recent governance instruments refuse to permit for high-stakes systems. The European Union's AI Act requires that high-risk AI be designed so that it can be effectively overseen by natural persons who can intervene or halt it, and it \blue{requires those persons to be enabled to remain aware of the tendency to over-rely on automated output, which the Act names as automation bias~\cite{euaiact2024}}.

This matters beyond assigning blame after the fact. Accountability is also forward-looking: it is the expectation of having to answer that disciplines a builder's choices in advance. A human who knows they will be asked to justify a deployment behaves differently from a system optimised only for a proxy metric. Governance frameworks encode this by making a named person accountable for an AI system's risks rather than the system itself~\cite{nistrmf2023}. Guardrails, checklists, and stated values do not survive contact with a process that has no one inside it who can be held to them.

\blue{A 2025 incident illustrates both halves of the hazard in miniature. During a publicised trial of an agentic coding platform, the agent deleted a live production database despite an explicitly declared code freeze; the vendor's chief executive described the event as unacceptable and one that ``should never be possible'', and the remedial measures were architectural rather than exhortatory---automatic separation of development from production data, improved rollback and a planning-only mode in which the agent cannot act~\cite{fortune2025replit}. Two things are notable. The action was irreversible as far as the agent was concerned and was recovered, in the event, from a platform snapshot, so the boundary that mattered was reversibility, not the instruction the agent had been given. And the party who answered publicly was a person. Both observations recur in the framework of Section~\ref{sec:framework}.}

\FloatBarrier
\section{Alternatives, Counter-Arguments and the Residual Human Contribution}\label{sec:alternatives}
\blue{The strongest objection to the argument so far is that it establishes a need for \emph{independent} assurance, not for \emph{human} assurance. This section takes the objection seriously in each of its forms, concedes what should be conceded, and identifies what is left.}

\subsection{``Humans share a substrate too''}\label{sec:humansubstrate}
\blue{People are trained on a common curriculum, read the same textbooks and copy the same idioms; Knight and Leveson's result is precisely that human teams working independently make correlated mistakes~\cite{knight1986}, and Eckhardt and Lee's model attributes it to inputs that are hard for everyone~\cite{eckhardt1985}. Multiple experts nonetheless help, and the same technique---ensembles of differently trained models---is used to improve machine judgement. Does the argument not cut equally against human oversight?}

\blue{It would, if the argument were that humans are independent of one another or of the problem. It is not. The claim concerns the covariance in~\eqref{eq:cov} between the \emph{verifier's} difficulty function and the \emph{builder's}. A human reviewer and a generative builder are trained by different processes on different data toward different objectives, and there is evidence that their failure geometries differ: adversarial perturbations that reliably defeat learned classifiers measurably influence human observers only under severe time limits, and then far short of the near-total misclassification seen between models~\cite{elsayed2018}, and the problem-space evasions that pass a malware classifier are constructed to preserve the malicious functionality that a human analyst inspects~\cite{pierazzi2020}. Cross-substrate correlation is thus expected to be lower than within-substrate correlation, which is what the argument needs. But the reviewer's analogy carries a lesson the original version of this argument did not draw: whether a given human is in fact independent of a given machine is an empirical matter, subject to the same measurement as any other pair (Section~\ref{sec:measure}), and Section~\ref{sec:humansbiased} shows that it is easily destroyed by the way the review is organised. The analogy also supports, rather than undermines, the use of heterogeneous machine ensembles, to which we now turn.}

\subsection{Heterogeneous model ensembles}\label{sec:ensembles}
\blue{An ensemble of models drawn from different vendors and lineages is the most direct technical answer to correlated failure, and the evidence is that it works in part. Ron et al.\ found that three-version majority voting over agent-generated implementations reduced mean observed failures from roughly 387 to 131~\cite{ron2026}; Nogueira et al.\ found that combining models lowered failure correlation, though the reliability gain was less than half of what independence would predict and depended far more on which models were combined than on language or prompting~\cite{nogueira2026}; Littlewood and Miller's theory allows that well-chosen methodological diversity can do better still~\cite{littlewood1989}. The framework of Section~\ref{sec:framework} therefore requires heterogeneity between builder and verifier as a matter of course.}

\blue{Three limits follow from the analysis of Section~\ref{sec:defs}. The first is that vendor labels are weak evidence of independence: vendors train on overlapping public corpora (D1), distil or fine-tune from common bases (D2), and evaluate against common suites (D3), and Tihanyi et al.'s nine models from several vendors exhibited broadly the same coding errors~\cite{tihanyi2025}. Independence must be measured by $\rho$, and re-measured whenever a model is updated. The second is that every member of a learned ensemble remains a learned verifier: it is exposed to the transferable evasions and shared-corpus poisoning of Section~\ref{sec:adversary}~\cite{demontis2019,carlini2024}, and Goel et al.'s finding that errors correlate more strongly as capability rises suggests that the most capable ensembles may be the least diverse~\cite{goel2025}. The third is dimension D6: an ensemble of any size still judges the artefact against a specification it did not write and cannot question. Ensembles restore independence for fault classes within the specification's reach. They do not restore it for the specification itself.}

\subsection{Deterministic tools and formal verification}\label{sec:deterministic}
\blue{Static analysers, type systems, sanitisers, fuzzers with crash oracles and proof assistants are not drawn from the generative substrate at all, and for the properties they cover they supply exactly the independence the closed loop lacks. A memory-safety sanitiser does not share a language model's prior about buffer arithmetic; a proof that an operating-system kernel refines its specification, as in seL4, is independent of any intuition the kernel's authors had~\cite{klein2009}. The framework accordingly treats the availability of such an oracle for the property at stake---the criterion we call \emph{verifiability}---as a principal reason to permit higher autonomy, and it prefers deterministic and formal oracles to learned ones wherever they exist.}

\blue{Their limits are well understood and are the same ones the seL4 team stated: a proof is relative to its specification and to assumptions about the compiler, the hardware and the initialisation code~\cite{klein2009}; an analyser covers the fault classes it was written for and is silent on authorisation logic, business-rule errors and everything else that is a flaw only relative to intent; a fuzzer finds what crashes, not what quietly leaks. Coverage, not independence, is their weakness. The consequence for the argument is that deterministic oracles push the residual human requirement toward the specification level rather than eliminating it: the more of the artefact-level checking that can be handed to mechanisms which do not share \B, the more clearly the remaining question is whether the specification is right.}

\subsection{Machine speed}\label{sec:speed}
\blue{Some security decisions must be taken faster than a person can be consulted---rate-limiting a credential-stuffing burst, sinkholing a volumetric attack, quarantining a host that is exfiltrating data. Requiring per-action approval for these would be the theatre that Section~\ref{sec:accountability} warned against, and would in practice be removed. The answer is not to exclude humans from these decisions but to move the human decision upstream. A pre-authorised \emph{action envelope}, set by an accountable person, defines the class of actions the machine may take at speed and the bounds within which it may take them; the defining property of an action inside the envelope is that it is reversible on a timescale the organisation can tolerate. Actions inside the envelope run at machine speed with post-hoc audit; actions outside it queue for approval or, if they cannot wait, are simply not available to the machine. The agentic-platform remedy described in Section~\ref{sec:accountability}---separating environments and guaranteeing rollback---is a minimal envelope of this kind~\cite{fortune2025replit}. Speed is therefore a reason to design bounded autonomy carefully, not a reason to abandon oversight.}

\subsection{Humans are biased too, and human independence is not given}\label{sec:humansbiased}
\blue{The evidence in Sections~\ref{sec:roof} and~\ref{sec:accountability} against human reliability---over-confident developers~\cite{perry2023}, complacent monitors~\cite{parasuraman2010}, deskilled operators~\cite{bainbridge1983}---is not an embarrassment for the argument; it is part of it. Treating the human as a naturally independent oracle is the weakest element of the conventional case for oversight. The correct statement is that a human's independence from the machine is a property of the pair that erodes with exposure and must be protected by design. Perry et al.'s participants were not independent of the assistant: they read its output and adopted its assumptions, so their $t$ became correlated with its $b$~\cite{perry2023}. Parasuraman and Manzey's experts lost their vigilance because the automation was reliably right, and recovered it when reliability varied~\cite{parasuraman2010}. Bu\c{c}inca et al.\ showed that cognitive forcing functions---designs that require the person to form a judgement before the machine's answer is shown---reduce over-reliance~\cite{bucinca2021}. Each of these findings is a design constraint: judgement before verdict, variable and seeded exposure so that vigilance is exercised, currency maintained by unaided practice, and reviewer independence from the builder in the ordinary organisational sense (here the IEEE 1012 proxies retain their value, because they were designed for people~\cite{ieee1012}). Section~\ref{sec:measure} gives the conditions and the indicators.}

\subsection{What remains for people, and under what conditions}\label{sec:residual}
\blue{Once heterogeneous ensembles, deterministic tools and bounded autonomy have been given their due, three functions remain that none of them supplies. We state each as a design argument, with its scope, rather than as a claim about human nature.}

\blue{\textbf{Specification authority.} Every machine oracle---learned, deterministic or formal---judges an artefact against a stated property. The requirement that the property was meant to capture belongs to whoever the system is for, and that party is a person or an organisation of persons. When a flaw is a requirement error (the specification permits the harm, or omits the case), no artefact-level oracle can detect it, and only the owner of the requirement is positioned to adjudicate it. This is not a claim that the owner is infallible; it is a claim about who is entitled to decide what the system ought to do, and it locates a human at the point where specifications are written and changed. The concentration of common-mode failures at hard or ambiguous specification clauses in the coding-agent replication of Knight and Leveson~\cite{ron2026} is the empirical face of this point.}

\blue{\textbf{Accountability.} Under the governance instruments in force, responsibility for the risks of an AI system attaches to a named natural or legal person, not to the system~\cite{nistrmf2023,euaiact2024}; the responsibility gap is a reason for this allocation, not an argument against it~\cite{matthias2004}. The scope is legal and organisational: a jurisdiction could in principle choose otherwise, and the argument would then have to be made afresh. Within present arrangements, accountability requires a person whose authority over the lifecycle is real, which in turn requires the information, competence and time set out in Section~\ref{sec:measure}.}

\blue{\textbf{Last-resort halt.} The adversarial analysis of Section~\ref{sec:adversary} showed that an attacker who has poisoned or evaded the substrate has compromised every check that runs on it. A halt authority that is exercised from outside the substrate is therefore not redundant with in-loop checks; it is the only control that survives their compromise. The authority need not be a person's finger on a switch---an independent deterministic watchdog can trigger it---but the decision to trust the watchdog, to set its thresholds and to act on what it reports is one that someone must own, for the reasons given under accountability.}

\blue{None of these three functions requires a person to approve individual actions. They require a person who owns the specification, who answers for the lifecycle, and who can stop it. How much more is required at any given control point depends on the consequence, reversibility, time-criticality, verifiability and adversarial exposure of the action in question. Making that dependence explicit is the purpose of the framework.}

\section{An Operational Framework for Bounded Autonomy}\label{sec:framework}
\blue{The framework has three parts: a scale of autonomy levels, a set of distinct human roles, and decision criteria that assign a maximum permissible level to each class of action in each lifecycle phase. It is deliberately compatible with the levels-of-automation model of Parasuraman, Sheridan and Wickens~\cite{parasuraman2000}, restricted to the decision and action-implementation stages that matter for security actions.}

\subsection{Autonomy levels}
\blue{Table~\ref{tab:levels} defines five levels. The critical distinction is between L2, at which a person approves each action (in the loop), and L3, at which a person defines an envelope in advance and supervises its execution (on the loop). Most of the disputes about ``human oversight'' are disputes about which of these two is meant, and the framework's central claim is that the answer differs by action and can be decided by explicit criteria.}

\begin{table}[!htbp]
\centering\small
{\tblue
\caption{Autonomy levels for security-lifecycle actions, adapted from the decision-and-action stages of the Parasuraman--Sheridan--Wickens model~\cite{parasuraman2000}.}
\label{tab:levels}
\begin{tabularx}{\textwidth}{@{}L{1.0cm} L{3.4cm} X L{3.6cm}@{}}
\toprule
\textbf{Level} & \textbf{Name} & \textbf{Machine and human behaviour} & \textbf{Human role engaged} \\
\midrule
L0 & Manual & Person performs the action; machine may supply information. & Operator \\
L1 & Advisory & Machine proposes; person decides and acts. Machine output is one input among others. & Operator \\
L2 & Approved autonomy (in the loop) & Machine prepares and, on explicit per-action approval, executes. Approval must be informed (evidence, provenance, uncertainty shown) and formed before the machine's recommendation is seen. & Operator; Supervisor sets what requires approval \\
L3 & Bounded autonomy (on the loop) & Machine executes autonomously within a pre-authorised envelope of reversible, bounded-consequence actions; person can veto or halt in real time and audits a sample after the fact. & Supervisor; Accountable owner sets envelope \\
L4 & Full autonomy (out of the loop) & Machine executes; people are informed after the fact or not at all. Permitted only where the criteria of Table~\ref{tab:rules} are all satisfied. & Accountable owner only \\
\bottomrule
\end{tabularx}}
\end{table}

\subsection{Three human roles}
\blue{Discussions of ``human oversight'' commonly treat the human as a single role. Three roles are in fact distinct in authority, timescale and competence, and conflating them is the source of much confusion about what oversight demands.}
\begin{itemize}
\item \blue{\textbf{Operator} (in the loop). Approves or performs individual actions at L0--L2. Needs task competence, the information listed in Table~\ref{tab:oversight}, and a time budget adequate to the decision. An operator who cannot meet these conditions is not oversight and should not be counted as such.}
\item \blue{\textbf{Supervisor} (on the loop). Defines and tunes action envelopes within policy, monitors L3 execution, exercises veto and halt, and audits samples. Needs situational awareness across the pipeline rather than per-item depth, and the vigilance supports of Section~\ref{sec:measure}.}
\item \blue{\textbf{Accountable owner} (governance). Owns the specification and the requirements it encodes, sets policy on which actions may reach L3 and L4, accepts residual risk, and is the named party who answers for outcomes at every level including L4. Must be organisationally independent of the team that builds or operates the models, in the IEEE 1012 sense~\cite{ieee1012}. This role, and not the operator, is where the three functions of Section~\ref{sec:residual} principally live.}
\end{itemize}

\subsection{Decision criteria}
\blue{Five properties of an action determine the maximum level at which it may be automated. \emph{Consequence} (C) is the severity of harm if the action is wrong. \emph{Reversibility} (R) is whether, and how quickly, the action can be undone; it is the single most important criterion, because reversibility converts an error into a cost. \emph{Time-criticality} (T) is the ratio of the decision window to the time a competent person needs to review; when the window is shorter, per-action approval is not available and the choice is between L3 and not automating at all. \emph{Verifiability} (V) is whether an oracle independent of the acting model exists for the property at stake and, if that oracle is learned, whether its measured $\rho$ against the actor is acceptably close to one; deterministic and formal oracles score highest. \emph{Adversarial exposure} (A) is whether the action's inputs can be shaped by an adversary, and whether the action touches the substrate itself (training data, model weights, detection thresholds, the verifier). Table~\ref{tab:rules} gives the assignment rule.}

\begin{table}[!htbp]
\centering\small
{\tblue
\caption{Assignment of the maximum permissible autonomy level from the five criteria. Thresholds for ``low'' and ``high'' are policy decisions to be fixed by the accountable owner and recorded; the structure of the rule is what the framework prescribes.}
\label{tab:rules}
\begin{tabularx}{\textwidth}{@{}L{1.6cm} X@{}}
\toprule
\textbf{Max level} & \textbf{Conditions} \\
\midrule
L4 & R high (undo is automatic and fast) \emph{and} C low \emph{and} V high (independent, preferably deterministic oracle with measured $\rho\approx 1$) \emph{and} A low. Example: enrichment and de-duplication of alerts; fuzzing in an isolated harness. \\
L3 & (R high \emph{or} V high) \emph{and} T high (decision window shorter than human review), \emph{or} any action that meets the L4 conditions except one. Envelope defined by the accountable owner; mandatory post-hoc audit sampling; real-time veto. Example: blocking a source with automatic expiry; deploying a patch that has passed deterministic regression tests to a canary population with automatic rollback. \\
L2 & C high \emph{or} R low, \emph{and} T low enough for informed review. Approval formed before the machine's recommendation is shown. Example: merging generated code that touches authentication, cryptography or authorisation; isolating a production system where no automatic rollback exists. \\
L1 & V low \emph{and} (C high \emph{or} A high). The machine informs; a person decides. Example: interpreting anomalous behaviour in a control system whose model of ``normal'' the adversary may have shaped. \\
Governance (always human) & Changes to the specification or requirements; changes to an action envelope; acceptance of residual risk; retraining or replacing a model used in any role (A high by definition); disabling a control; incident sign-off. \\
\bottomrule
\end{tabularx}}
\end{table}

\subsection{Applying the framework across the lifecycle}
\blue{Table~\ref{tab:phases} applies the rule to representative actions in each phase. Two features deserve comment. First, the same phase contains actions at every level: ``should the tester be autonomous?'' is the wrong question, and ``should autonomous exploit generation be permitted against isolated targets, and against live ones?'' has two different answers. Second, the requirement of heterogeneity between builder and verifier (D1--D3 of Table~\ref{tab:dimensions}, measured by $\rho$) is a precondition for any verifier to raise V at all; a learned verifier drawn from the builder's substrate contributes nothing to V regardless of how many of them are used.}

\begin{table}[!htbp]
\centering\small
{\tblue
\caption{Representative actions by lifecycle phase, with the maximum permissible level under Table~\ref{tab:rules} and the control that makes the level acceptable.}
\label{tab:phases}
\begin{tabularx}{\textwidth}{@{}L{1.2cm} L{5.2cm} L{1.5cm} X@{}}
\toprule
\textbf{Phase} & \textbf{Action} & \textbf{Max} & \textbf{Making it acceptable} \\
\midrule
Build & Generate code for a component with no security-sensitive surface, gated by deterministic tests and a heterogeneous reviewer & L3 & Measured $\rho$ between generator and reviewer; deterministic gates; revertible merge \\
Build & Generate or modify authentication, cryptographic or authorisation code & L2 & Independent human review formed before seeing the generator's rationale; heterogeneous machine review as a second input \\
Build & Change the specification, threat model or acceptance criteria & Gov. & Requirement owner decides; machine may draft, not decide \\
Defend & Enrich, de-duplicate and rank alerts & L4 & No action taken on the environment; audit of ranking drift \\
Defend & Block a source or throttle a session with automatic expiry & L3 & Envelope with expiry and rate limits; post-hoc sampling \\
Defend & Isolate a production host & L3$^{a}$ & Rollback tested; consequence bounded by envelope \\
Defend & Retrain or update a detection model on newly collected data & Gov.\ then L2 & Poisoning canaries; hold-out evaluation by an independent party; owner sign-off \\
Test & Fuzz or statically analyse in an isolated harness & L4 & Isolation verified; findings queue for triage \\
Test & Generate and run exploits against an isolated replica & L3 & Envelope: replica only; network egress blocked; halt authority \\
Test & Exploit or scan against live production targets & L2 & Explicit, scoped authorisation per target; time-boxed; halt authority present \\
Test & Deploy an autonomously generated patch to production & L3$^{b}$ & Canary rollout; independent (non-substrate) verifier of the patched property \\
\bottomrule
\end{tabularx}
\par\smallskip\footnotesize $^{a}$L3 where automatic restoration exists and has been tested; otherwise L2.\quad $^{b}$L3 where a deterministic regression suite and automatic rollback exist; otherwise L2.}
\end{table}

\FloatBarrier
\subsection{When higher autonomy is justified, and the trade-offs}\label{sec:tradeoffs}
\blue{The framework is not a presumption in favour of human involvement. Higher autonomy is justified, and should be adopted, when an action is reversible, when an oracle independent of the actor exists and has been measured, when the decision window is shorter than competent review, when volume would otherwise degrade review into rubber-stamping, and when delay itself benefits the adversary. Alert triage, fuzzing, isolated exploitation and canary patching all meet these conditions today, and the AI Cyber Challenge results show what is gained by permitting them~\cite{darpaaixcc2025}.}

\blue{The trade-offs run in both directions, and misplaced oversight has costs of its own. Per-action approval at high volume produces exactly the complacency Parasuraman and Manzey measured~\cite{parasuraman2010} and yields a lower effective catch rate than sampled audit of a well-bounded L3 envelope; requiring people to approve what they cannot competently evaluate produces the theatre that regulators explicitly reject~\cite{euaiact2024}; and removing people from routine work removes the practice that keeps their intervention competent~\cite{bainbridge1983}. The framework's answer to the last of these is that competence maintenance is a control to be scheduled, not a by-product to be hoped for (Table~\ref{tab:oversight}). Scalability is bought by widening envelopes, not by thinning approval, and each widening is a governance decision recorded against the person who made it.}

\FloatBarrier
\section{Measuring Independence and Oversight Effectiveness}\label{sec:measure}
\blue{A framework whose requirements cannot be audited is a set of intentions. This section specifies how the two properties on which the framework depends---failure independence between roles, and the effectiveness of human oversight---can be measured, and states the central claims as hypotheses.}

\subsection{Measuring independence}\label{sec:measure-indep}
\blue{The coincident-failure ratio $\rho$ of~\eqref{eq:rho} is estimated on a \emph{probe set}: a corpus of situations with known flaws spanning the fault classes of concern (for code, the relevant weakness enumerations; for detection, attack techniques), some produced by the builder under test and some seeded independently of it, together with matched flaw-free controls. The builder's emission rate is measured on its own outputs; the verifier's miss rate is measured on the independently seeded items, so that the two marginals are not confounded; the escape rate is the fraction of all builder outputs that are flawed and pass the verifier, which equals the emission rate multiplied by the verifier's miss rate on the builder's own flawed outputs. Reporting $\rho$ per fault class with confidence intervals, rather than a single number, matters because a verifier can be independent of a builder on memory-safety flaws and correlated with it on authorisation logic. For pairs of classifiers, the chance-adjusted agreement-on-mistakes statistic of Goel et al.\ is an alternative with a known baseline~\cite{goel2025}; for generated implementations, the behavioural-diversity methodology of Nogueira et al.\ applies directly~\cite{nogueira2026}. Because model updates change $\rho$, it should be re-measured on every change of model, vendor or fine-tune in either role, and the record kept with the envelope decisions it justified. What counts as an acceptable $\rho$ is a policy decision for the accountable owner; the framework requires only that the decision be made against a measurement.}

\subsection{Conditions for competent, timely and effective human review}
\blue{Table~\ref{tab:oversight} states the conditions under which a human review is capable of contributing independence, together with an indicator for each. The indicators are chosen so that failure is visible: an operator who never overrides a machine whose measured error rate is non-zero is not reviewing, whatever the process document says.}

\begin{table}[!htbp]
\centering\small
{\tblue
\caption{Conditions for effective human oversight and the indicator that shows whether each is met.}
\label{tab:oversight}
\begin{tabularx}{\textwidth}{@{}L{2.6cm} X L{5.2cm}@{}}
\toprule
\textbf{Condition} & \textbf{Requirement} & \textbf{Indicator} \\
\midrule
Competence & Reviewer can perform the task unaided; currency maintained by scheduled unaided practice~\cite{bainbridge1983} & Unaided catch rate on seeded faults, measured periodically \\
Time & Decision window exceeds the time a competent reviewer needs; otherwise the action belongs in an L3 envelope, not an L2 queue & Median review time versus window; queue depth \\
Information & Evidence, provenance, the machine's uncertainty and the applicable envelope are available at the point of decision & Audit of decision records for missing fields \\
Judgement before verdict & Reviewer records a judgement before the machine's recommendation is revealed~\cite{bucinca2021} & Proportion of decisions with pre-verdict record; disagreement rate \\
Vigilance & Reviewer is exposed to variable reliability and to seeded faults so that checking is exercised~\cite{parasuraman2010} & Seeded-fault catch rate over time; override rate versus measured machine error rate \\
Organisational independence & Reviewer and accountable owner are distinct from the team that builds or operates the models~\cite{ieee1012} & Reporting lines; conflict-of-interest register \\
Halt authority & A tested halt exists and can be exercised within the consequence horizon & Time-to-halt in exercises; halt drills logged \\
\bottomrule
\end{tabularx}}
\end{table}

\FloatBarrier
\subsection{Hypotheses and how they would be refuted}\label{sec:hypotheses}
\blue{The argument rests on four propositions. We state them so that they can be tested, and we say what would count against each.}
\begin{description}[leftmargin=1.2em,style=nextline]
\item[\blue{H1 (within-substrate correlation).}] \blue{For a builder and a learned verifier drawn from the same substrate, $\rho>1$ for security-relevant fault classes, and the excess is largest for flaws that are semantic rather than syntactic. Refuted if $\rho$ is statistically indistinguishable from~1 across classes for same-substrate pairs. Existing evidence for pairs of generated implementations~\cite{ron2026,nogueira2026} and for model-as-judge~\cite{goel2025,panickssery2024} is consistent with H1; the three-role version has not been tested.}
\item[\blue{H2 (proxies versus measurement).}] \blue{Heterogeneity reduces $\rho$, but organisational labels (different vendor) predict $\rho$ poorly relative to measured overlap on D1--D3. Refuted if vendor identity alone accounts for most of the variance in $\rho$ across model pairs.}
\item[\blue{H3 (erosion of human independence).}] \blue{A human reviewer's $\rho$ against a machine builder is higher when the machine's verdict is shown before the reviewer's judgement is recorded than when it is shown after. Refuted if the order has no measurable effect. Perry et al.\ and Bu\c{c}inca et al.\ supply indirect support~\cite{perry2023,bucinca2021}.}
\item[\blue{H4 (misplaced oversight).}] \blue{At high action volume and fixed machine error rate, sampled audit of a bounded L3 envelope achieves a higher effective catch rate than per-action L2 approval. Refuted if per-action approval maintains its catch rate as volume rises. Parasuraman and Manzey's results predict H4~\cite{parasuraman2010}.}
\end{description}
\blue{A protocol to test H1 and H2 follows directly from Section~\ref{sec:measure-indep}: fix a probe set, vary the builder and verifier across same-substrate, cross-vendor and deterministic pairings, and estimate $\rho$ per class. H3 and H4 are experiments in the tradition of Perry et al.\ and Parasuraman and Manzey, with $\rho$ and catch rate as the outcome measures. We regard running this protocol as the natural next step, and we have designed the framework so that an organisation adopting it would generate the data as a by-product of its own audits.}

\FloatBarrier
\section{Discussion and Limitations}\label{sec:discussion}
\blue{This is a conceptual and analytical article. It contributes definitions, a model, a framework and metrics, and it draws on published measurements, but it reports no new experiment. Its central empirical proposition is therefore a hypothesis, and we have tried to keep the distinction between hypothesis and result visible throughout, including in the title, which speaks of measurable independence rather than of measured independence.}

\blue{Several further limitations should be stated. Two of the most directly relevant studies~\cite{ron2026,nogueira2026} and two others~\cite{hubinger2024,shojaee2025} are preprints, and the argument would weaken if their peer-reviewed forms differed materially; we have relied on them for direction rather than for magnitudes. The three residual human functions of Section~\ref{sec:residual} are design arguments whose scope is bounded by present governance arrangements and by the current state of machine reasoning; either could change, and the framework is intended to be re-applied, not fixed. The thresholds in Table~\ref{tab:rules} are deliberately left to policy, which is a strength for adoption and a weakness for comparability across organisations. The framework addresses the software security lifecycle; extending it to cyber-physical systems, where reversibility is often unavailable by nature, would require a separate treatment of consequence. Finally, the metric $\rho$ is only as good as the probe set, and an adversary who learns the probe set can optimise against it; probe sets should be treated as sensitive and rotated.}

\blue{The practical implication for organisations is nonetheless immediate. Autonomy can be widened today for the large class of reversible, verifiable, bounded actions, and doing so will improve both speed and, by relieving reviewers of volume they cannot competently handle, the quality of oversight where it remains. The precondition is that independence between the roles be measured rather than inferred from a vendor's name, and that a person own the specification, the envelopes and the halt.}

\section{Conclusion}\label{sec:conclusion}
AI now builds, defends, and tests, and it will do more of each. \blue{The temptation is to read that trajectory as ending in a lifecycle that runs itself, and the customary rejoinder is that a human must remain in the loop. Neither is quite right. A closed loop of models certifying models loses the independence that makes verification informative, and it does so in a way that the organisational proxies of existing standards cannot detect; but much of that independence can be restored by heterogeneous and, above all, deterministic oracles, provided their independence is measured rather than assumed. What cannot be restored by such means is narrower than a blanket requirement for oversight and more durable than one: authority over what the system is for, accountability for what it does, and the power to stop it when every internal check may have been turned. The three hats can be worn by machines. The head that answers for them cannot, under any governance arrangement now in force, and the task is not to keep that head everywhere in the lifecycle but to place it where measurement shows independence is otherwise absent, where consequences cannot be undone, and where an answer will be demanded.}

\FloatBarrier
\section*{Declarations}
\noindent\textbf{Conflict of interest.} \blue{The author declares no conflict of interest.}\\
\noindent\textbf{Funding.} This research received no specific grant from any funding agency.\\
\noindent\textbf{Data availability.} No datasets were generated or analysed. The metric definitions and the measurement protocol are given in full in Sections~\ref{sec:defs} and~\ref{sec:measure}.\\
\noindent\textbf{Author contributions.} \blue{M.C.G.\ conceived the argument, conducted the analysis and wrote the manuscript.}\\
\noindent\textbf{Use of generative artificial intelligence.} Generative-AI tools were used to assist with language editing and reference checking; the author reviewed and takes full responsibility for the content.

\bibliographystyle{unsrtnat}
\bibliography{references}

\end{document}